\documentclass[]{elsart}
\usepackage{amsmath}
\usepackage{amssymb}
\usepackage{graphicx}

\def\issue(#1,#2,#3){#1 (#3) #2} 

\def\opcit(#1){ {\em op. cit.}, #1}

\def\ARNPS(#1,#2,#3){Ann.\ Rev.\ Nucl.\ Part.\ Sci.\ \issue(#1,#2,#3)}
\def\CPC(#1,#2,#3){Comp.\ Phys.\ Comm.\ \issue(#1,#2,#3)}
\def\CIP(#1,#2,#3){Comput.\ Phys.\ \issue(#1,#2,#3)}
\def\EPJC(#1,#2,#3){Eur.\ Phys.\ J.\ C\ \issue(#1,#2,#3)}
\def\IEEETNS(#1,#2,#3){IEEE Trans.\ Nucl.\ Sci.\ \issue(#1,#2,#3)}
\def\NP(#1,#2,#3){Nucl.\ Phys.\ \issue(#1,#2,#3)}
\def\NIM(#1,#2,#3){ Nucl.\ Instrum.\ and Meth.\ \issue(#1,#2,#3)}
\def\PL(#1,#2,#3){Phys.\ Lett.\ \issue(#1,#2,#3)}
\def\PRD(#1,#2,#3){Phys.\ Rev.\ D \issue(#1,#2,#3)}
\def\PRL(#1,#2,#3){Phys.\ Rev.\ Lett.\ \issue(#1,#2,#3)}
\def\SJNP(#1,#2,#3){Sov.\ J. Nucl.\ Phys.\ \issue(#1,#2,#3)}
\def\ZPC(#1,#2,#3){Z.\ Phys.\ C \issue(#1,#2,#3)}
\def\PAN(#1,#2,#3){Phys.\ Atom.\ Nucl.\ \issue(#1,#2,#3)}
\def\FBS(#1,#2,#3){Few\ Body\ Syst.\ \issue(#1,#2,#3)}

\begin{document}

\begin{frontmatter}

\title{Search for a strongly decaying neutral charmed pentaquark}


\collaboration{The~FOCUS~Collaboration}\footnotemark
\author[ucd]{J.~M.~Link}
\author[ucd]{P.~M.~Yager}
\author[cbpf]{J.~C.~Anjos}
\author[cbpf]{I.~Bediaga}
\author[cbpf]{C.~Castromonte}
\author[cbpf]{A.~A.~Machado}
\author[cbpf]{J.~Magnin}
\author[cbpf]{A.~Massafferri}
\author[cbpf]{J.~M.~de~Miranda}
\author[cbpf]{I.~M.~Pepe}
% ...

\address[ucd]{University of California, Davis, CA 95616}
\address[cbpf]{Centro Brasileiro de Pesquisas F\'\i sicas, Rio de Janeiro, RJ, Brazil}
% ...

\footnotetext{See \textrm{http://www-focus.fnal.gov/authors.html} for additional author information.}

\begin{abstract}
We present a search for a charmed pentaquark decaying strongly to $D^{(*)-}p$.  Finding no
evidence for such a state, we set limits on the cross section times branching ratio
relative to $D^{*-}$ and $D^-$ under particular assumptions about the production mechanism.
\end{abstract}


\end{frontmatter}

\section{Introduction}

At the dawn of the QCD era, Jaffe proposed the existence of bound 
(mass below threshold for strong decay) multiquark states including
$Q\overline{Q}q\overline{q}$ states and the $H$ dihyperon~\cite{jaffe} based on calculations
using the bag model~\cite{bag}.  Calculations a decade later~\cite{lipkin,gignoux,karl} 
indicated that a $S=-1$ charmed
pentaquark $(\overline{c}sqqq)$ is more likely to be bound than either the $H$ dihyperon 
or a $S=0$ charmed pentaquark $(\overline{c}qqqq)$ due to the lack of a quark exchange 
process in the lowest decay channel, $D_s^-N$~\footnote{Charged conjugate states are
implied unless explicitly stated otherwise}.  
Additional calculations using bag models and group theory assuming a one-gluon exchange 
interaction suggested that $S=0,-1$ charmed pentaquarks were 
probably unbound due to SU(3) symmetry breaking and the finite mass of the charm
quark~\cite{group,bag_1,bag_2}.  Calculations using an instanton model indicated a nearly 
bound $S=-1$ charmed pentaquark~\cite{instanton} while calculations based on Goldstone 
boson exchange predicted unbounded $S=-1$ charm pentaquarks but allowed bounded
or unbounded $S=0$ charm pentaquarks~\cite{goldstone_1,goldstone_2}. 
Finally, the Skyrme (chiral soliton) models predicted deeply bound 
charmed pentaquark states with $S=0,-1$~\cite{skyrme_1,skyrme_2,skyrme_3}. 
Searches in 1998 and 1999 found no evidence for a weakly decaying charmed $S=-1$ pentaquark
from $\pi^-N$ interactions~\cite{e791_penta}.


The pentaquark field underwent a transformation between January 2003 and March 2004, when no less than 
ten independent pentaquark observations at a mass around $1540\;\textrm{MeV}\!/c^2$ were 
reported~\cite{allobs}, with a presumed quark content
of $(\overline{s}uudd)$.  This triggered a rash of theoretical explanations, including predictions of 
charmed pentaquark states based on various quark models~\cite{penta_preh1_qm}, 
Skyrme models~\cite{penta_preh1_skyrme}, and lattice QCD~\cite{penta_preh1_lattice} which predicted
$\Theta_c^0(\overline{c}uudd)$ masses of $2600$--$3000$~MeV/$c^2$, $2700$~MeV/$c^2$, and $3450$~MeV/$c^2$,
respectively, compared to the $DN$ threshold of $\sim\! 2805$~MeV/$c^2$.
In March 2004, the H1 Collaboration reported evidence for a $S=0$ charmed pentaquark state 
decaying to $D^{*-}p$ at a mass of $3099 \pm 3 \pm 5$~MeV$/c^2$ and a statistical significance of
$5.4\:\sigma$ or $6.2\:\sigma$, depending on how the significance is calculated~\cite{h1}.  This 
precipitated additional theoretical effort into pentaquark states with a heavy quark~\cite{penta_charm_2}.

This letter presents a search for the $\Theta_c^0(\overline{c}uudd)$ pentaquark candidate.  
We search using the same decay
mode as H1 $(\Theta_c^0\!\to\!D^{*-}p)$ and the other obvious decay mode, 
$\Theta_c^0\!\to\!D^-p$.  Since the $D^{*+}$ statistics and data quality of the FOCUS experiment are 
much better than the observing experiment, and the production mechanism is similar, we should
be able to confirm or refute the existence of the purported state.

\section{Event reconstruction and selection}

The FOCUS experiment recorded data during the 1996--7 fixed-target run at Fermilab.  
A photon beam obtained from bremsstrahlung of 300~GeV electrons and positrons impinged
on a set of BeO targets.  Four sets of silicon strip detectors, each with three views, 
were located downstream of the targets for vertexing and track finding.  For most of 
the run, two pairs of silicon strips were also interleaved with the target segments for more
precise vertexing~\cite{tsilicon}.  Charged particles were tracked and momentum analyzed as they passed
through one or two dipole magnets and three to five sets of multiwire proportional chambers 
with four views each.  Three multicell threshold \v{C}erenkov counters, two electromagnetic
calorimeters, and two muon detectors provided particle identification.  A hadronic trigger 
passed 6 billion events for reconstruction.  The average photon energy of reconstructed charm
events is 175~GeV.

A candidate driven vertexing algorithm is used to reconstruct charm.  In the case of
$D^0\!\to\!K^-\pi^+,K^-\pi^+\pi^-\pi^+$ $(D^+\!\to\!K^-\pi^+\pi^+)$, the charged tracks are required to 
verticize with CL $>$ 2\% (1\%) 
with the correct summed charge.  The momentum and vertex 
location are used as a ``seed'' track to find the 
production vertex, which must have CL $>$ 1\%.   
The $D^{*+}$ candidate is obtained from the 
decay $D^{*+}\!\to\!D^0\pi_s^+$.  The soft pion must
be consistent with originating from the production vertex and the track is refit using the production
vertex as an extra constraint.  

The selection criteria were chosen to maximize $S/\sqrt{B}$.  The
signal, $S$, comes from a Monte Carlo simulation.  Since the production characteristics of the pentaquark
are unknown, we use \textsc{Pythia} to generate charm events with a $\Xi_c^0$.  The
$\Xi_c^0$ is produced at a mass of $2.47$~GeV/$c^2$ but is promoted to a mass of 
$3.1$~GeV/$c^2$ before decaying to the decay mode of interest (with zero lifetime).  
The background, $B$, is obtained from the wrong sign data $(D^{(*)+}p)$ over the 
entire mass range of study (threshold to $4$~GeV/$c^2$).  This is an unbiased method for determining
the selection criteria.

\begin{figure}
\centerline{\includegraphics[width=1.82in,height=2.2in]{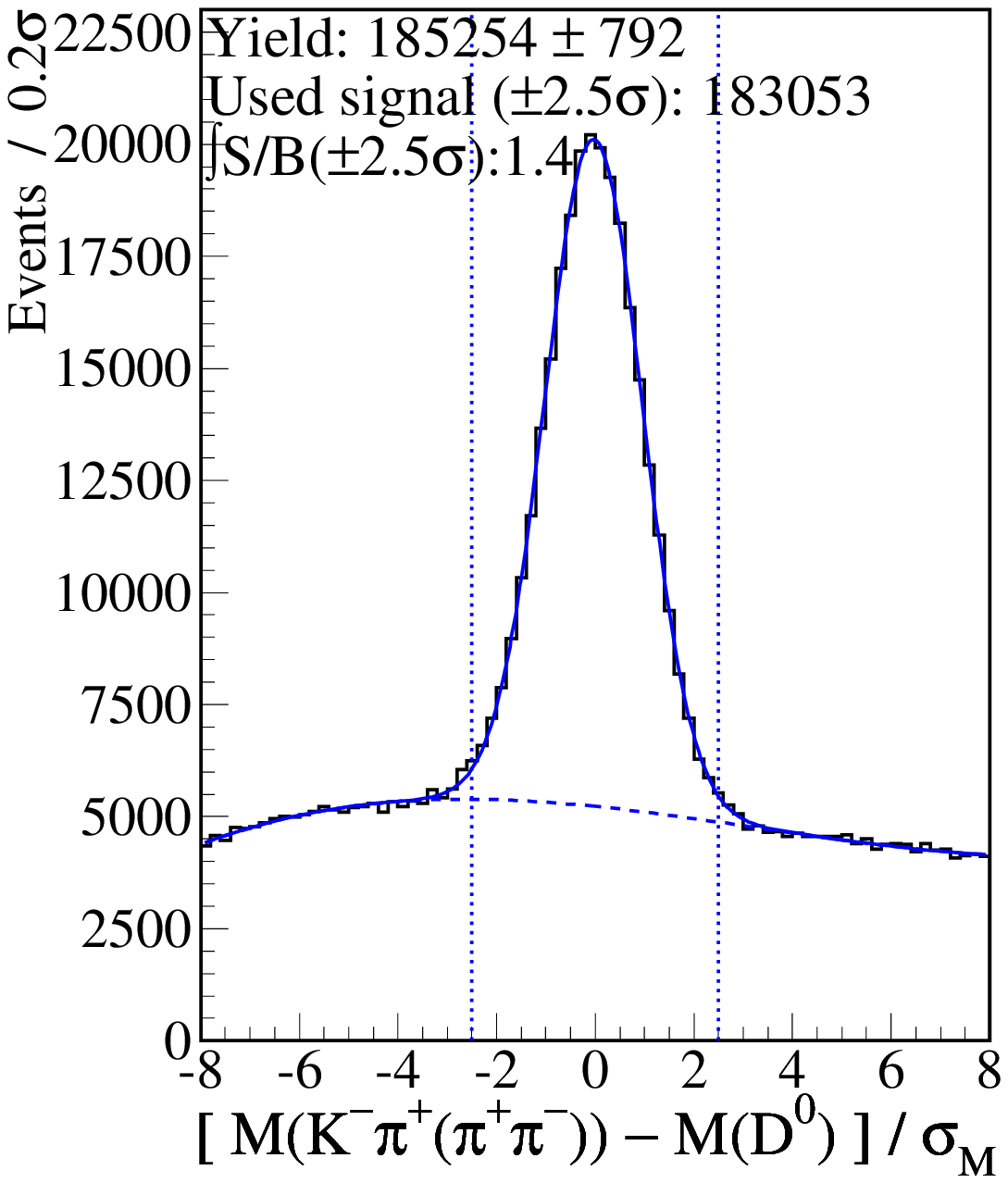}\hspace{-3pt}
\includegraphics[width=1.82in,height=2.2in]{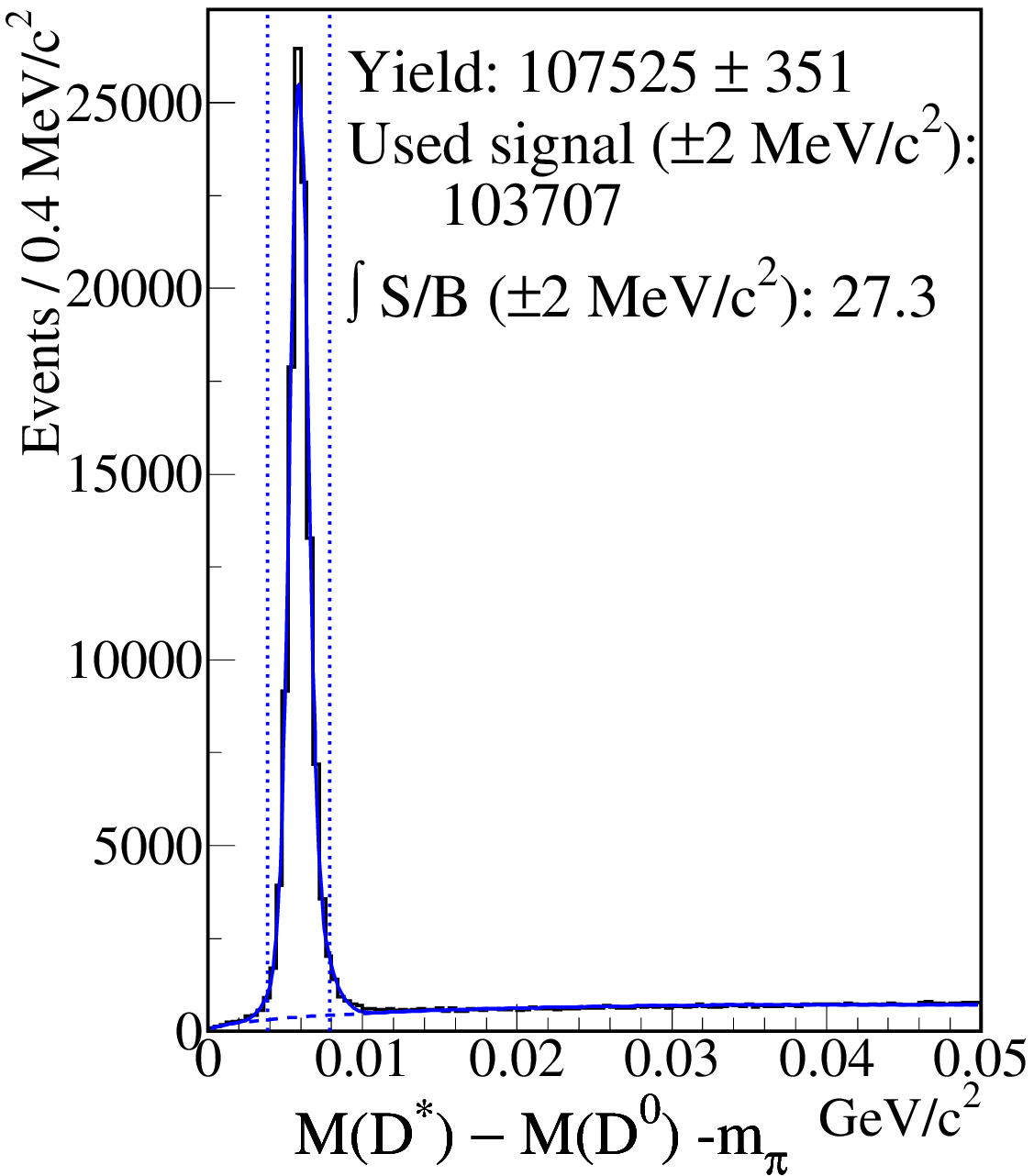}\hspace{-5pt}
\includegraphics[width=1.82in,height=2.2in]{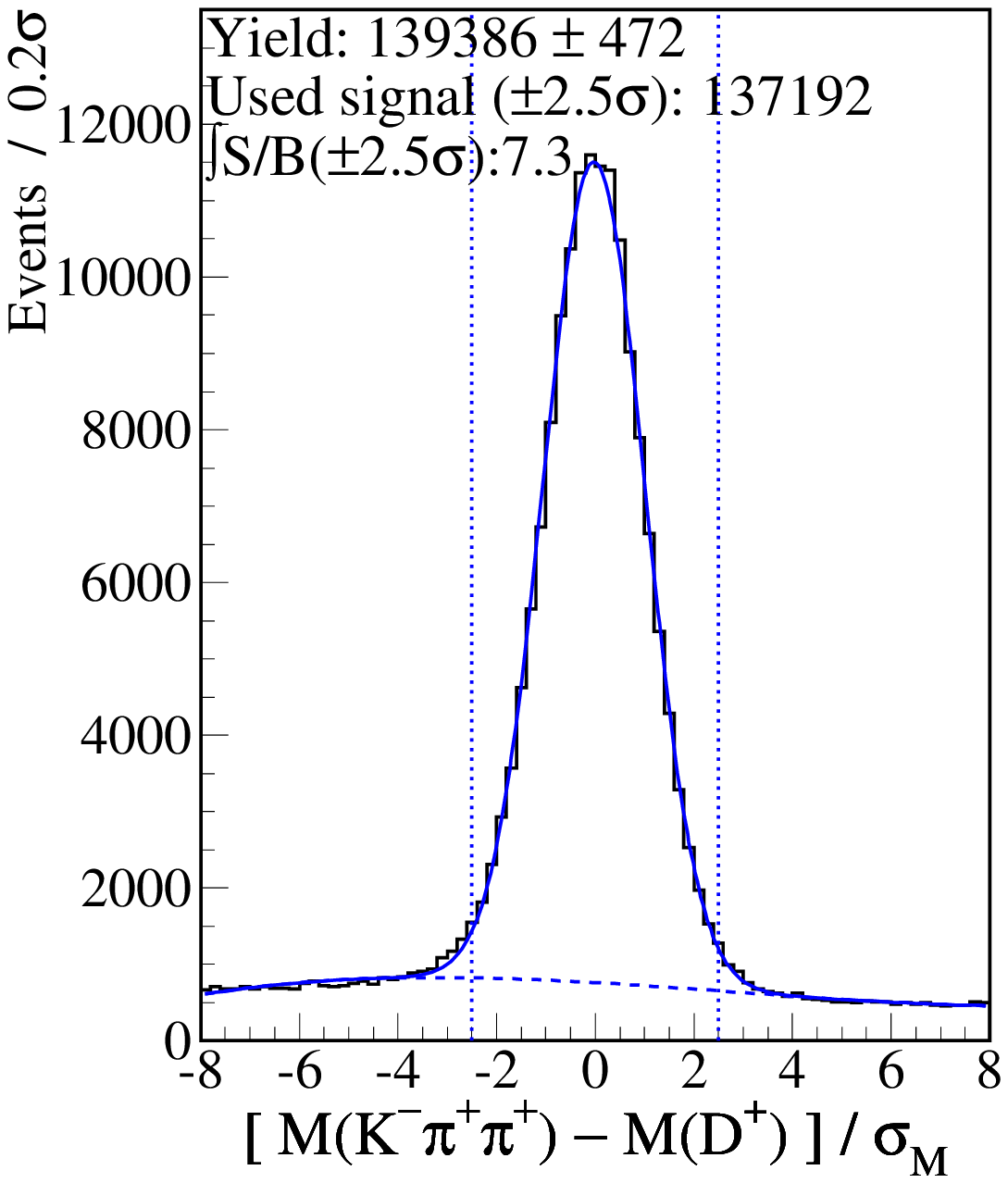}}
\caption{The normalized mass plots of $D^0\!\rightarrow\!K^-\pi^+$ and $D^0\!\rightarrow\!K^-\pi^+\pi^-\pi^+$ candidates 
(left) and $D^+\!\rightarrow\!K^-\pi^+\pi^+$ candidates (right) are fit with a single Gaussian for the signal and 
a quadratic polynomial for the background.  The energy release plot for $D^{*+}$ candidates of 
$D^0$ events (middle) is fit with a double Gaussian for the signal plus a threshold function 
$\alpha Q^{1/2} + \beta Q^{3/2}$ for the background.  Events inside the vertical lines are selected for analysis.}
\label{fig:charmsig}
\end{figure}

Separating charm from hadronic background is primarily accomplished by requiring the decay
vertex be distinct from the production vertex.  A cut of $\ell/\sigma_\ell > 2, 4, 6$ is applied
to $D^0\!\to\!K^-\pi^+$, $D^0\!\to\!K^-\pi^+\pi^-\pi^+$, $D^+\!\to\!K^-\pi^+\pi^+$ where $\ell$ 
is the distance between the two vertices and $\sigma_\ell$ is 
the calculated uncertainty ($\langle$$\sigma_\ell$$\rangle$ $\sim$ 500 $\mu$m).
Since hadronic reinteractions can fake a decay, requiring the decay vertex be located
outside of target material reduces background.  The out-of-material significance $\sigma_\textrm{out}$ is
positive (negative) for a vertex outside (inside) material.  We require 
$2\,\ell/\sigma_\ell + \max{\left(-2,\sigma_\textrm{out}\right)} > 4, 8, 12$ for 
$D^0\!\to\!K^-\pi^+$, $D^0\!\to\!K^-\pi^+\pi^-\pi^+$, $D^+\!\to\!K^-\pi^+\pi^+$.  To ensure the $D^0$ decay tracks
do not originate from the production vertex, a cut is made on the change in production vertex confidence
level when any $D^0$ decay track is added to the vertex ($\Delta \textrm{CL}_\textrm{pri}$).  
We require $\Delta \textrm{CL}_\textrm{pri} < 30\%\;(0)$ 
for $D^0$ $(D^+)$ decays.  Also, the transverse
momentum of the $D$ candidate with respect to the line-of-flight obtained from the vertices must be less
than $0.7$, $0.4$, and $0.3$~GeV/$c$ for $D^0\!\to\!K^-\pi^+$, $D^0\!\to\!K^-\pi^+\pi^-\pi^+$, and
$D^+\!\to\!K^-\pi^+\pi^+$, respectively.
Since the signal contains a charged kaon,
information from the three \v{C}erenkov counters effectively suppresses backgrounds.  The 
\v{C}erenkov identification algorithm~\cite{citadl} returns negative log-likelihood (times two) values 
$\mathcal{W}_i(j)$ for track $j$ and hypothesis $i\in\left\{e,\pi,K,p\right\}$.  In practice,
differences in log-likelihoods between hypotheses are used such as 
$\Delta\mathcal{W}_{\pi K} \equiv \mathcal{W}_\pi-\mathcal{W}_K$.  For the $D^{*-}$ candidates, we require 
$\Delta\mathcal{W}_{\pi K}(K)>0.5$,
$\mathcal{W}_\textrm{min}(\pi)-\mathcal{W}_\pi(\pi) > -5$, and
$\mathcal{W}_\textrm{min}(\pi_s)-\mathcal{W}_\pi(\pi_s) > -5$ where 
$\mathcal{W}_\textrm{min} \equiv \min{(\mathcal{W}_{i\in\{e,\pi,K,p\}})}$.   
For the $D^+$ candidates, we require 
$\Delta\mathcal{W}_{\pi K}(K)>2$ and
$\mathcal{W}_\textrm{min}(\pi)-\mathcal{W}_\pi(\pi) > -2$.
The remaining selection criteria are very efficient and mildly suppress some
backgrounds.
For the $D^0\!\to\!K^-\pi^+$ decay, the momentum asymmetry, $\left|\frac{p(K)-p(\pi)}{p(K)+p(\pi)}\right|$, must be
less than $\frac{p(D)+130\:\textrm{GeV}/c}{200\:\textrm{GeV}/c}$.  The summed $p_T^2$
of $D$ daughters with respect to the $D$ momentum vector must be greater than $0.15\:\textrm{GeV}^2/c^2$.
These last two cuts favor a decay of a heavy particle over combinatoric background.
The charm signals obtained with these cuts are shown in Fig.~\ref{fig:charmsig}.  We select $D^0$ and $D^+$ 
candidates with an invariant mass within $2.5\:\sigma_M$ of the nominal mass where $\sigma_M$ is the calculated
error on the mass.  We select $D^{*+}$ candidates within 2~MeV/$c^2$ of the nominal energy release, 
$Q\equiv M(D^{*+})-M(D^0)-m_{\pi^+}$.  These values were also chosen in the same optimization procedure.
The proton candidate must be consistent with originating from the production vertex and must be strongly
favored to be a proton by the \v{C}erenkov systems; $\Delta\mathcal{W}_{\pi p}(p)>6$ and
$\Delta\mathcal{W}_{K p}(p) > 2$.  Again, these criteria were selected in the unbiased manner
described above.

\section{Pentaquark search results}

\begin{figure}
\centerline{\includegraphics[width=2.7in]{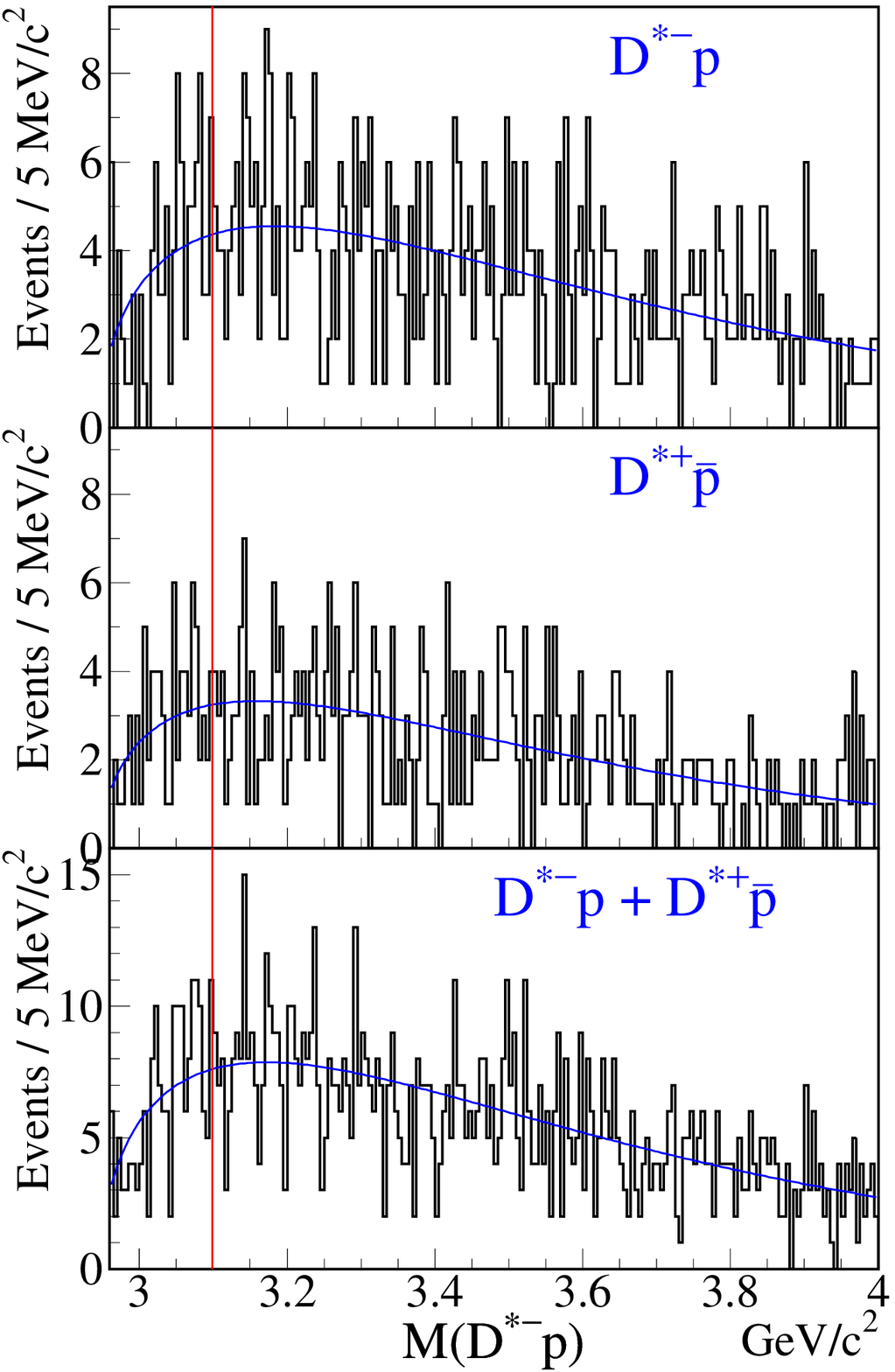}\hspace{-3pt}
\includegraphics[width=2.7in]{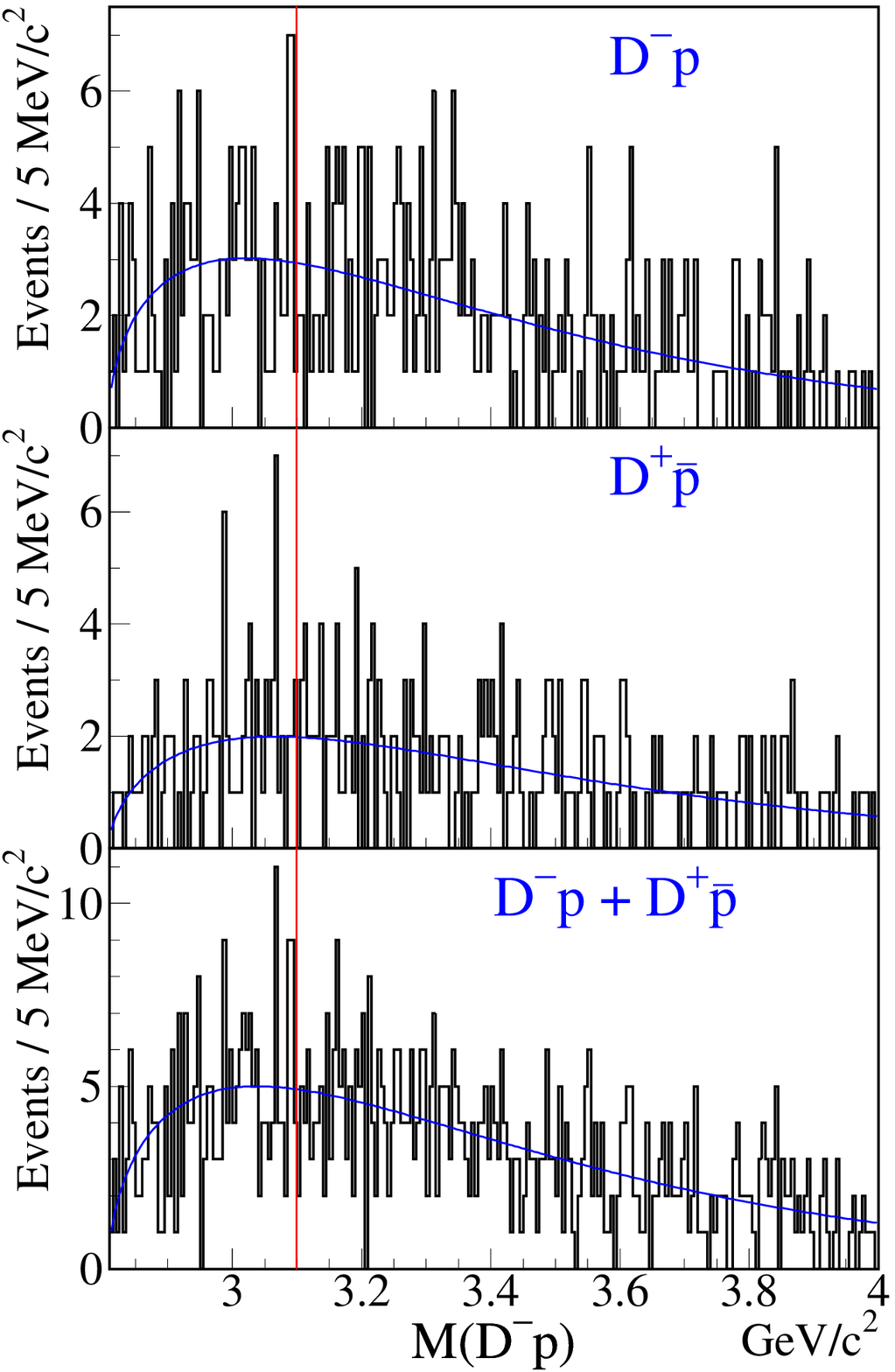}}
\caption{Invariant mass plots of $D^{*-}p$ candidates (left) and $D^-p$ candidates (right).  The top and 
middle plots show the $\overline{D}p$ and $D\overline{p}$ contributions, respectively.  The bottom
plot includes both charge conjugate states.  Plots are fit to a
background shape: $aQ^b \exp{(cQ)}$ where $Q \equiv M(\Theta_c)-M(D^{(*)-})-m_p$ is the energy release.}
\label{fig:pentac_bg}
\end{figure}

The $D^{*-}p$ and $D^-p$ invariant mass plots, separated by charge states and summed together, are shown in 
Fig.~\ref{fig:pentac_bg} with fitted background curves superimposed.  
The individual charge states show approximately an equal number of events in each state and so for the remainder
of the analysis we combine the charge conjugate states.
No evidence for a pentaquark at $3.1$~GeV/$c^2$ or at any mass less than $4$~GeV/$c^2$ is observed.
To set a limit on the yield we first need information about the width of the state.  H1 reports a measured 
Gaussian width of $\sigma_T = 12 \pm 3$~MeV/$c^2$ with an expected resolution of $\sigma_R = 7 \pm 2$~MeV/$c^2$.  
We define the ``natural'' width as $\sigma_N = \sqrt{\sigma_T^2 - \sigma_R^2}$.  Since $\sigma_T$ and $\sigma_R$ 
agree at better than 95\% CL, 
the lower limit on $\sigma_N$ is $0$.  
Using the above measurements to construct a $\chi^2$, we set 
$\chi^2 = 3.84$ to find the 95\% CL 
upper limit on $\sigma_N$.  The maximum value of $\sigma_N$ for $\chi^2 = 3.84$ 
occurs at $\sigma_N = 16.6$~MeV/$c^2$ with $\sigma_R = 6.1$~MeV/$c^2$ and $\sigma_T = 17.7$~MeV/$c^2$.  
We construct limits on yields under the two extreme assumptions about the pentaquark ``natural'' width, 
$\sigma_N = 0$ and $\sigma_N = 16.6$~MeV/$c^2$.  To a very good approximation, the upper limit increases 
monotonically with $\sigma_N$ so this provides the extreme range.  In both cases, 
$\sigma_N$ is added in quadrature to the experimental resolution to obtain the total width, $\sigma_T$.  
The experimental resolution increases linearly from 2~MeV/$c^2$ at threshold to 14 (13) MeV/$c^2$ at 
3.95~GeV/$c^2$ for $D^{*-}p$ ($D^-p$).

\begin{figure}
\centerline{\includegraphics[width=5.5in]{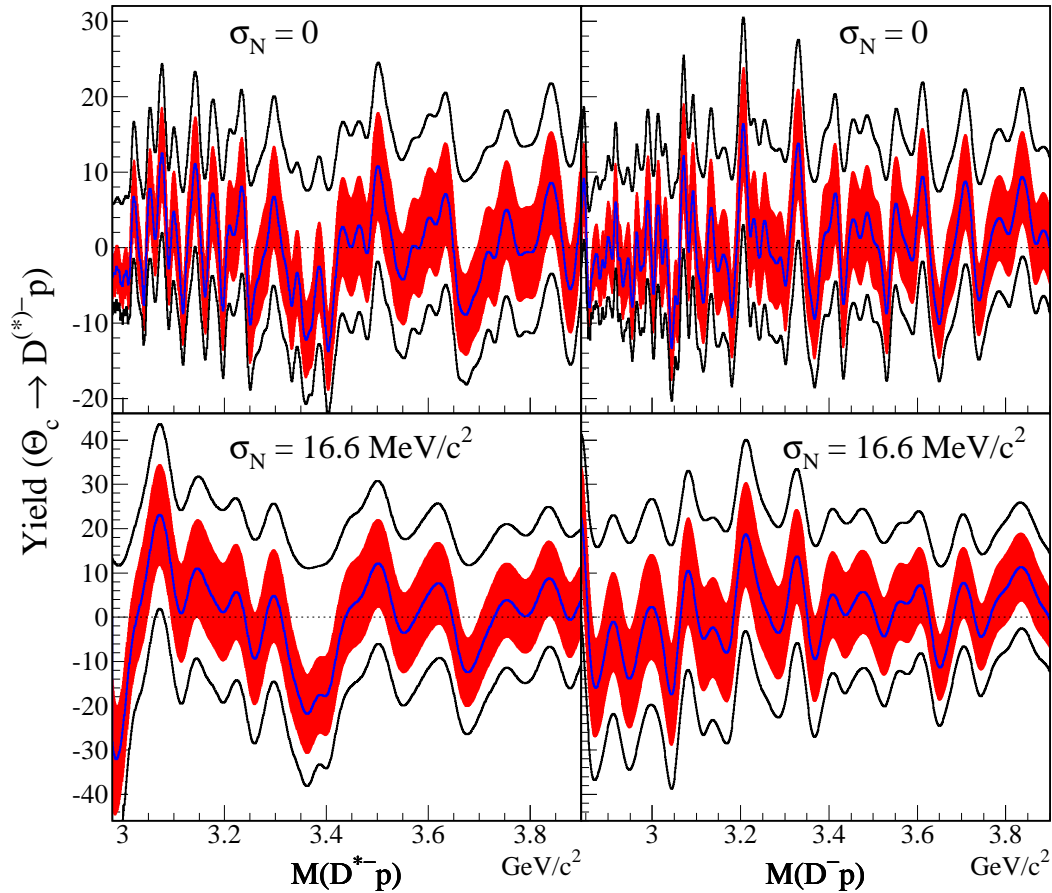}}
\caption{Charmed pentaquark yields and upper limits.  Left (right) plots show results for 
$\Theta_c\!\rightarrow\!D^{*-}p$ ($\Theta_c\!\rightarrow\!D^{-}p$).  Top (bottom) plots show
results for a natural width of $0$ ($16.6$~MeV/$c^2$).  The shaded region includes
the $1\:\sigma$ errors with the central value in the middle.  The outer curves show the upper and lower
limits.}
\label{fig:pentac_yld}
\end{figure}

The $D^-p$ $(D^{*-}p)$ mass plots are fit in 1~MeV/$c^2$ steps from $2.84$ $(2.98)$ to $3.9$~GeV/$c^2$.
At each step, the mass and width of the Gaussian signal are fixed while the signal yield and background 
parameters are fitted.  A binned log-likelihood fit using \textsc{Minuit}~\cite{minuit} is performed with 
a background shape given by $aQ^b \exp{(cQ)}$ where $Q$ is the energy release.  
The $1\:\sigma$ error is defined as the point where $\Delta \log{\mathcal{L}} = 0.50$ relative to 
the maximum $\log{\mathcal{L}}$, while continually adjusting the background parameters to maximize 
$\log{\mathcal{L}}$.  
The 95\% CL 
lower limit is defined similarly with $\Delta \log{\mathcal{L}} = 1.92$. 
Both are obtained using \textsc{Minos}~\cite{minuit}.  The 95\% CL 
upper limit is constructed as follows.  
The likelihood function $\mathcal{L}$ versus yield is determined by maximizing $\log{\mathcal{L}}$ for 
many different (fixed) yields, allowing background parameters to float.  The likelihood
function is integrated from a yield of $0$ to $\infty$.  The 95\% CL 
upper limit on the yield is defined 
as the point where 95\% 
of the total likelihood is between a yield of $0$ and the upper limit.
The fitted yield, 1-$\sigma$ errors, and 95\% CL 
limits are shown in Fig.~\ref{fig:pentac_yld}.

To compare with other experiments, the limits on yield must be converted to limits on production times 
(unknown) branching ratio.  We choose to normalize the $\Theta_c^0$ production 
cross section to the $D$ meson production cross section from which it is reconstructed.  
That is, we attempt to determine
$\sigma\left(\Theta_c^0\right) \cdot BR\left(\Theta_c^0\!\to\!D^-p\right) / \sigma\left(D^-\right)$ and 
$\sigma\left(\Theta_c^0\right) \cdot BR\left(\Theta_c^0\!\to\!D^{*-}p\right) / \sigma\left(D^{*-}\right)$.

\begin{figure}
\centerline{\includegraphics[width=5.5in]{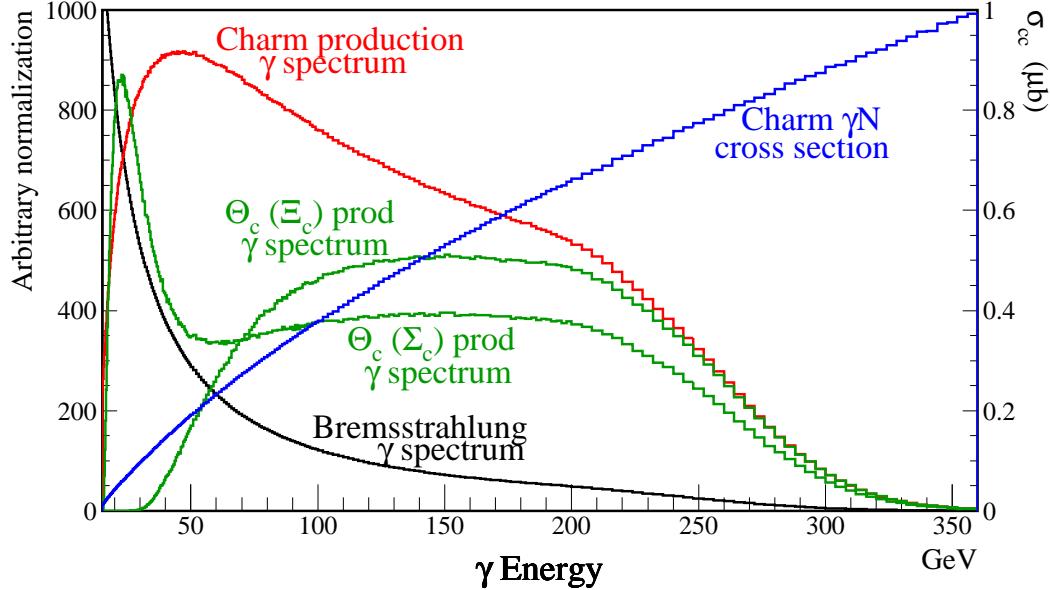}}
\caption{The initial bremsstrahlung spectrum peaks at low energies.  
Applying the rising charm cross section results in a harder photon spectrum for charm inducing photons.
The photons producing 3.1~GeV/$c^2$ $\Xi_c$ and $\Sigma_c$ particles are harder than generic charm 
inducing photons except for a low energy peak for $\Sigma_c$ production due to the charm quark combining
with quarks from the target nucleon.}
\label{fig:brem_charmxsec}
\end{figure}

\begin{figure}
\centerline{\includegraphics[width=5.5in]{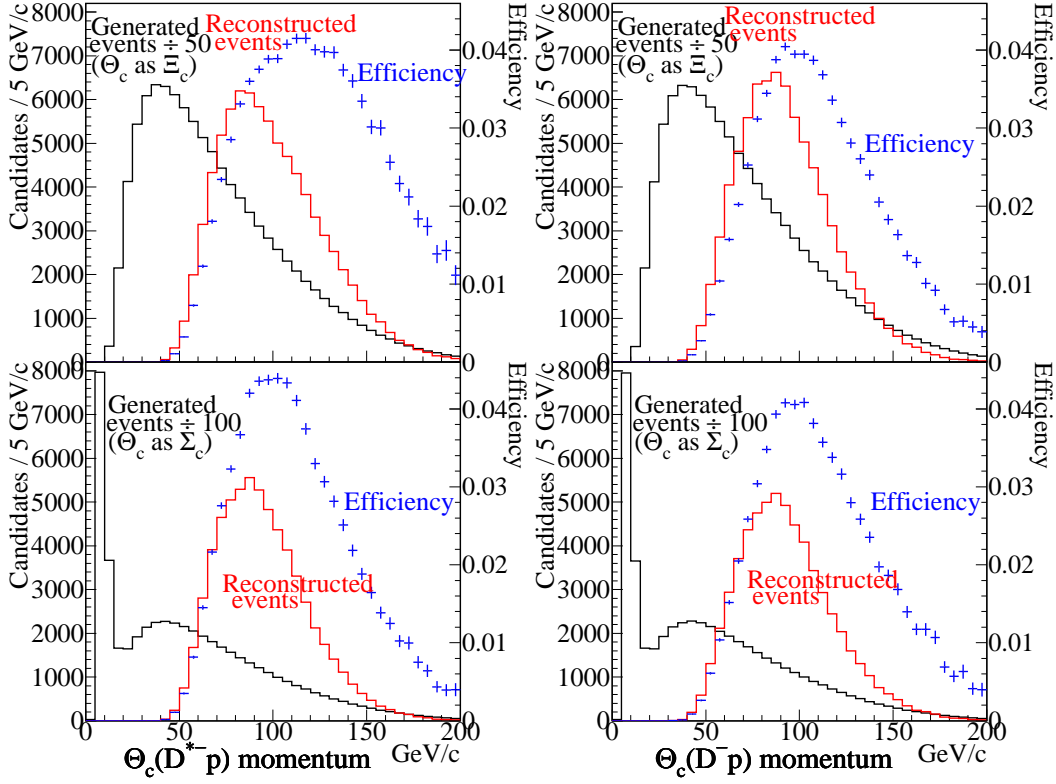}}
\caption{All four plots show generated and reconstructed events (histograms) and the calculated
efficiency (points with error bars) versus pentaquark momentum.  The top (bottom) plots are for 
pentaquarks produced as $\Xi_c^0$ $(\Sigma_c^+)$.  The left (right) plots are for the decay
$\Theta_c^0\!\rightarrow\!D^{*-}p$ $(\Theta_c^0\!\rightarrow\!D^-p)$.}
\label{fig:acc_vs_p}
\end{figure}

\begin{figure}
\centerline{\includegraphics[width=2.6in]{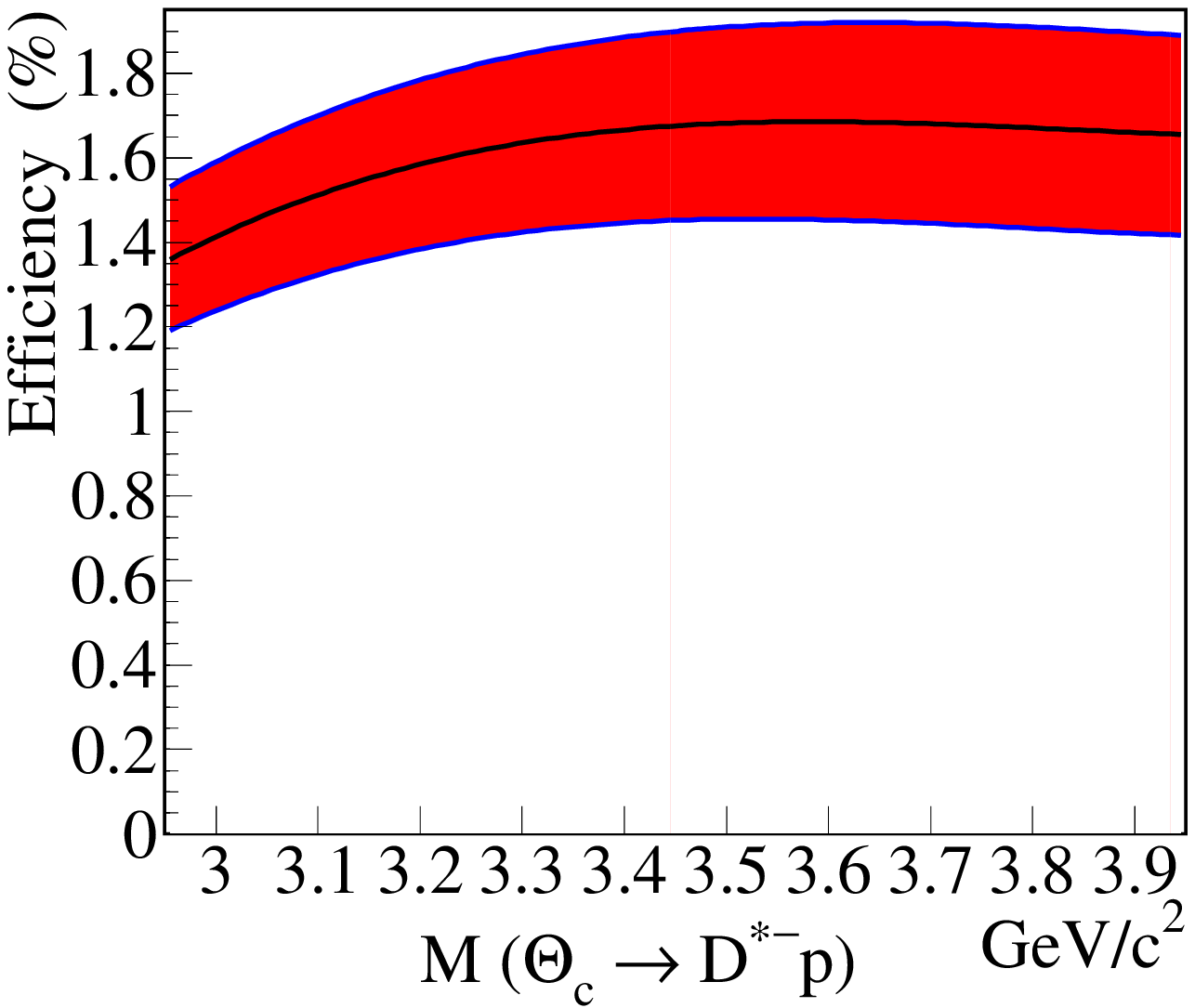}
\includegraphics[width=2.8in]{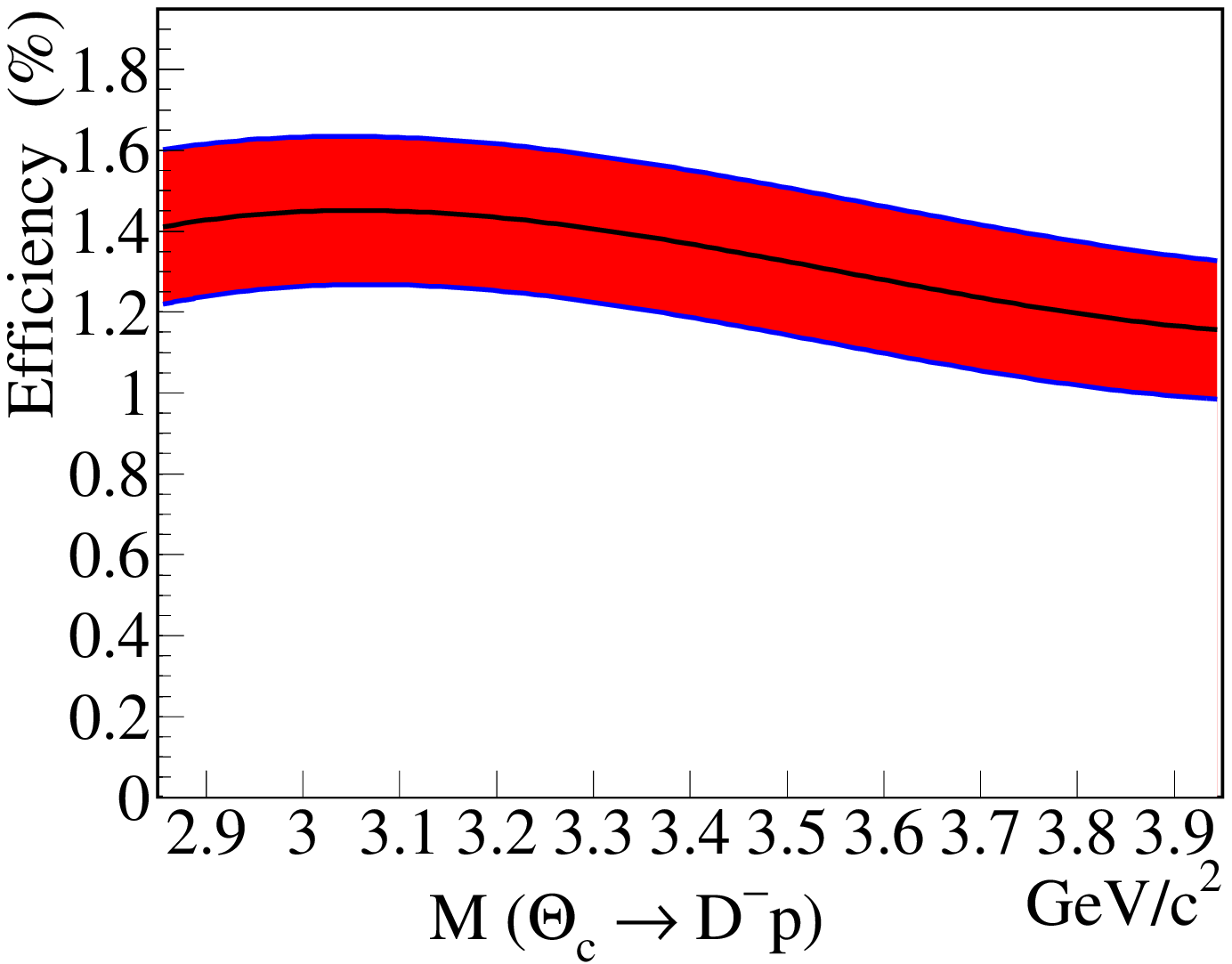}}
\caption{Efficiency (including acceptance) versus mass for reconstructing 
$\Theta_c^0\!\rightarrow\!D^{*-}p$ (left) and $\Theta_c^0\!\rightarrow\!D^-p$ (right).  
The top (bottom) curves are obtained from pentaquarks produced as $\Xi_c^0$ $(\Sigma_c^+)$ while the
middle curve is the average.}
\label{fig:acc_vs_m}
\end{figure}

The FOCUS detector is a forward spectrometer and therefore acceptance depends on
the produced particle momentum.  The production characteristics of the charmed pentaquark is by far the
largest source of systematic uncertainty in this analysis.  
We choose a particular production model to obtain limits and provide
sufficient information about the experiment for other interested parties to obtain limits based on other production 
models.  The production simulation begins with a library of $e^-$ and $e^+$ tracks obtained from a 
TURTLE simulation~\cite{turtle} of the Wideband beam line.  
From this library, an individual track is drawn and bremsstrahlung photons created by passage through 
a 20\% 
$X_0$ lead radiator.  The charm cross section is then applied to the bremsstrahlung photons using 
Monte Carlo rejection.  A photon which produces a charm event is
sent to the \textsc{Pythia} event generator~\cite{pythia}.\footnote{\textsc{Pythia} version 6.127 is used with
modifications to many parameters which provides a better match to the FOCUS charm data: 
MSTP(14):$\,20\!\to\!0$, PARP(2):$\,10\!\to\!3$, PMAS(4,1):$\,1.35\!\to\!1.6$, 
MSTJ(21):$\,2\!\to\!0$, PARJ(1):$\,0.1\!\to\!0.3$, PARJ(3):$\,0.4\!\to\!0.25$, MSTJ(12):$\,2\!\to\!4$, 
MSTP(92):$\,4\!\to\!2$, PARP(96):$\,3\!\to\!2$, PARP(97):$\,1\!\to\!2$, PARP(91):$\,0.44\!\to\!0.6$, 
PARP(93):$\,2\!\to\!3$, PARP(99):$\,0.44\!\to\!0.6$, PARP(100):$\,2\!\to\!3$, PARJ(21):$\,0.36\!\to\!0.425$,
MSTJ(13):$\,0\!\to\!1$, PARJ(41):$\,0.3\!\to\!0.25$, PARJ(42):$\,0.58\!\to\!0.7$, PARJ(45):$\,0.5\!\to\!0.75$,
PARJ(36):$\,2\!\to\!1$.}  Since \textsc{Pythia} does not produce charmed pentaquarks, another particle must be
chosen to represent the charmed pentaquark.  Charmed baryons are the natural choices since they have a 
single charm
quark like the charmed pentaquark and are closer to the correct number of total quarks than charmed mesons.
Since it is possible to adjust the mass of the chosen charmed baryon in \textsc{Pythia}, it is not necessary to
pick the charmed baryon with the highest mass.  Other than mass, the most important effect on the production
is the number of quarks a particle has in common with the initially interacting hadrons, due to
the nature of the \textsc{Pythia} string fragmentation model.
The $\Xi_c^0$ and $\Sigma_c^+$ particles are chosen to represent the extremes in the
production of a charmed pentaquark.  
Other than the charm quark, the $\Xi_c^0(csd)$ $(\Sigma_c^+(cud))$ 
can obtain at most 50\% (100\%) 
of the remaining quarks from the target nucleon valence quarks, 
while the $\Theta_c^0(\overline{c}uudd)$ can take 75\%.  
In all cases,
the charge conjugate particles must obtain all quarks from the vacuum.  
The mass of the particle chosen to represent the
pentaquark, $\Xi_c^0$ or $\Sigma_c^+$, is set to the appropriate value in \textsc{Pythia}, 
by setting PMAS(190,1) or PMAS(187,1), respectively.  This method differs from that used to generate the 
Monte Carlo sample for cut optimization since in this case the produced mass is changed to the appropriate 
value.
For a photon which passes the charm cross section,
the \textsc{Pythia} generator is run up to $1000$ times searching for a $\Xi_c^0$ $(\Sigma_c^+)$.  
This fails approximately 90\% 
of the time, especially for low energy photons.  
When this happens, a new charm producing photon is selected.  This changes the photon spectrum for 
pentaquark producing events.  
The initial bremsstrahlung spectrum, applied charm cross section, charm producing photon spectrum, 
and pentaquark producing photon spectra for $\Xi_c^0$ and $\Sigma_c^+$ are shown in Fig.~\ref{fig:brem_charmxsec}.  
After production, the $\Xi_c^0$ or $\Sigma_c^+$ is changed to the $\Theta_c^0$ with zero
lifetime and forced to decay in the mode of interest.  The momentum of the generated particles, the
momentum of the reconstructed particles, and the ratio of the two (efficiency versus momentum) are shown in 
Fig.~\ref{fig:acc_vs_p} for a pentaquark mass of $3.1$~GeV/$c^2$.  Since there are many low momentum 
pentaquarks from $\Sigma_c^+$ which are not reconstructed, the overall efficiency for pentaquarks produced as
$\Sigma_c^+$ is lower than for pentaquarks produced as $\Xi_c^0$.  The low generated momentum spike for 
pentaquarks produced as $\Sigma_c^+$ is a result of the produced charm quark combining with a $u$ and a $d$ quark
from the target.
The efficiency versus mass is shown in Fig.~\ref{fig:acc_vs_m} where the top (bottom) curve shows the efficiency
for pentaquarks produced as $\Xi_c^0$ $(\Sigma_c^+)$.  

The limits on $\Theta_c^0$ production
are made relative to $D^{(*)-}$ production.  Therefore, the efficiency for $D^{*-}$ and $D^-$ must be
determined.  \textsc{Pythia} is again used to model charm production.  In this case we have verified 
that \textsc{Pythia} correctly models charm production in the region over which we have acceptance.
This is less than 10\% 
of the total charm produced.  Note, however, that only the ratio of efficiencies
between $\Theta_c^0$ and $D^{(*)-}$ production need be determined.  Therefore, discrepancies between
\textsc{Pythia} and reality in the low momentum (unobserved) region are likely to cancel.  The global
efficiency for $D^{*-}$ $(D^-)$ reconstruction is $2.81\%$ $(3.05\%)$.

\begin{figure}
\centerline{\includegraphics[width=5.5in]{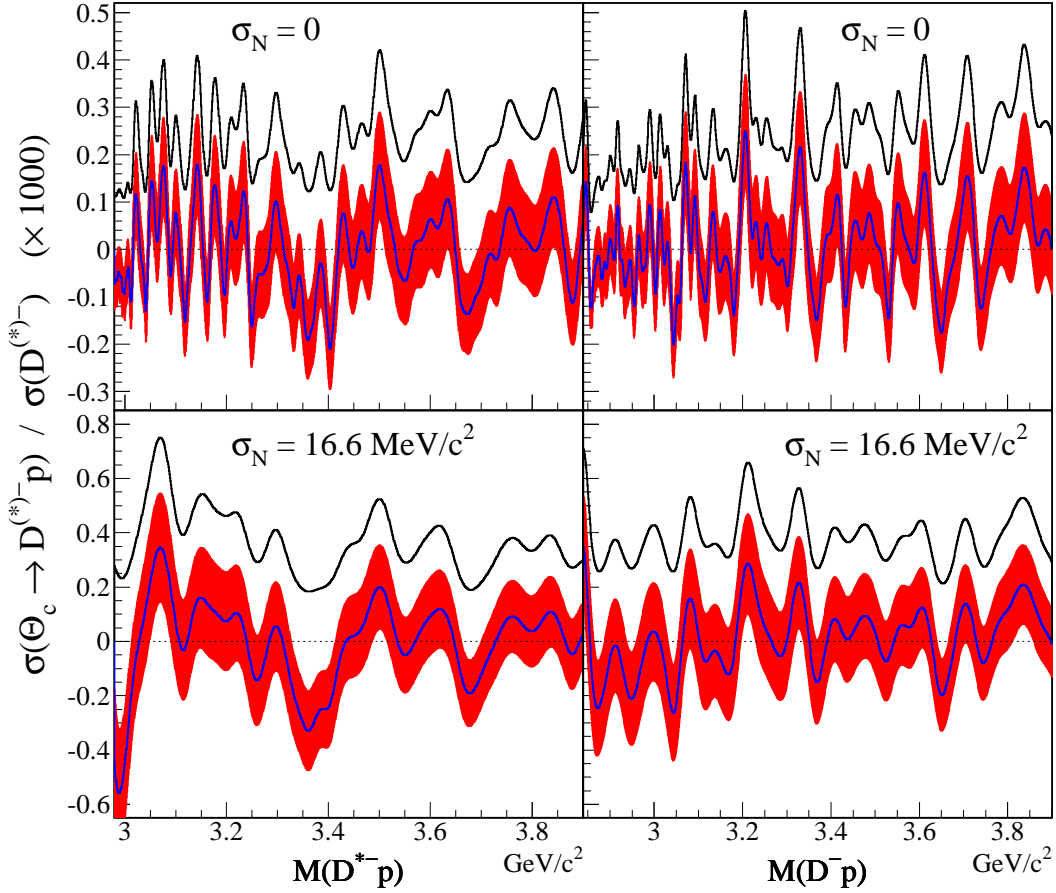}}
\caption{Cross section times branching ratio results relative to single $D$ production.  
Left (right) plots show results for 
$\Theta_c\!\rightarrow\!D^{*-}p$ ($\Theta_c\!\rightarrow\!D^{-}p$).  Top (bottom) plots show
results for a natural width of $0$ ($16.6$~MeV/$c^2$).  The shaded region encompasses
the $1\:\sigma$ errors with the central value in the middle.  The top curve shows the upper limit.
Both include systematic uncertainties.}
\label{fig:pentac_xsec}
\end{figure}

We attempt to determine:
\begin{equation}
\frac{\sigma\left(\Theta_c^0\!\rightarrow\! D\,p\right) \cdot \textrm{BR}\left(\Theta_c^0\!\rightarrow\! D\,p\right)}
{\sigma\left( D \right)} \;=\; \frac{Y_{\Theta_c^0}}{\epsilon_{\Theta_c^0}} \:\frac{\epsilon_D}{Y_D}
\end{equation}
where $D$ is either $D^{*-}$ or $D^-$.  The $D$ yield ($Y_D$) and pentaquark yield ($Y_{\Theta_c^0}$) are
obtained from Fig.~\ref{fig:charmsig} and Fig.~\ref{fig:pentac_yld}, respectively.
The uncertainty on the $D$ yield is less than 1\%,
insignificant compared to other uncertainties in the analysis.  
The remaining quantity is the ratio of efficiencies, 
$\epsilon_D/\epsilon_{\Theta_c^0}$.  Our uncertainty in $\epsilon_D$ is much less than $\epsilon_{\Theta_c^0}$
since the Monte Carlo simulation uses a tuned version of \textsc{Pythia} which accurately reproduces the
$D$ production for the region in which the FOCUS detector has acceptance.  The dominant source of systematic
uncertainty comes from a lack of knowledge of the pentaquark production characteristics which is reflected
in $\epsilon_{\Theta_c^0}$.  An estimate of this systematic uncertainty is obtained from Fig.~\ref{fig:acc_vs_m}
which shows the efficiency for $\Theta_c^0$ produced as $\Xi_c^0$ (top curve), $\Sigma_c^+$ (bottom curve),
and the average (middle curve).  We take the average as our central value for the efficiency $(\epsilon_0)$ 
and the difference between the average and top (or bottom) curve as the $1\:\sigma$ error on the efficiency
$(\sigma_{\epsilon_0})$.  To include the error we fit directly for the corrected yield 
$Y_{\Theta_c^0} / \epsilon_{\Theta_c^0}$, include the efficiency as a free parameter and use a modified
log-likelihood: 
$\log{\mathcal{L'}} = \log{\mathcal{L}} - 0.5\, (\epsilon_\textrm{fit} - \epsilon_0)^2 / \sigma_{\epsilon_0}^2$
where $\epsilon_\textrm{fit}$ is the fitted efficiency which is always equal to $\epsilon_0$ when 
$\log{\mathcal{L'}}$ is maximized.  The corrected yield, $1\:\sigma$ errors, and 95\%~CL 
upper limit is obtained
in the same manner as described previously for the uncorrected yield and is shown in Fig.~\ref{fig:pentac_xsec}.

\section{Conclusions}

\begin{figure}
\centerline{\includegraphics[width=5.5in]{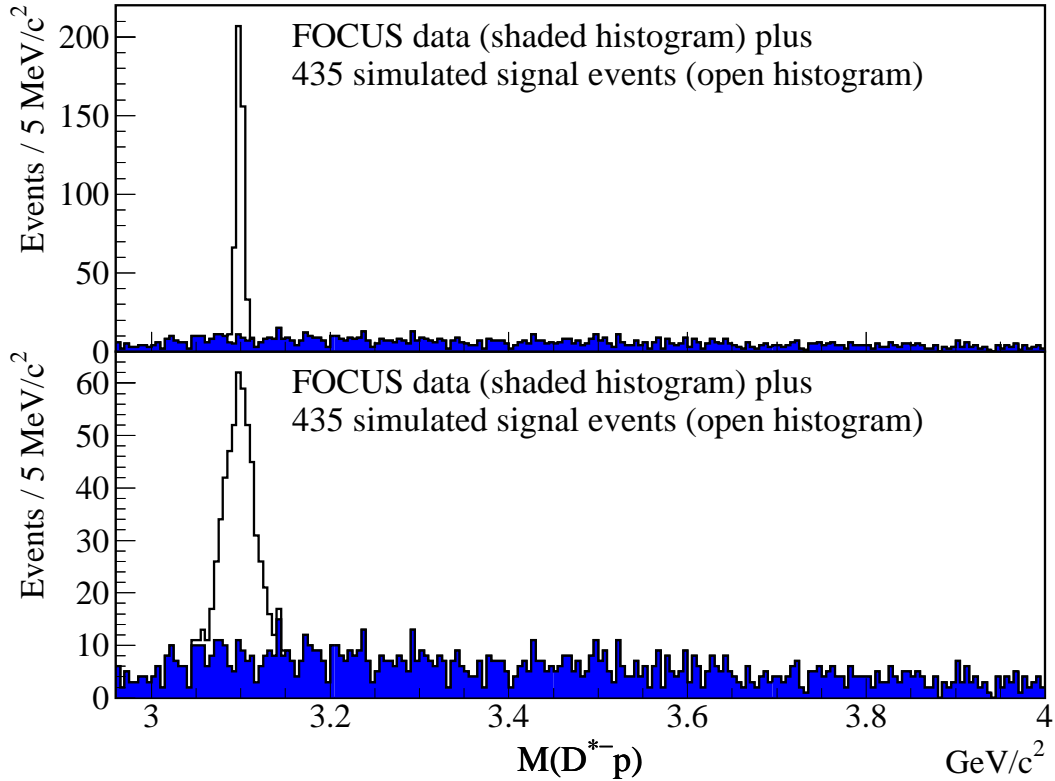}}
\caption{The FOCUS data (from the left plot of Fig~\ref{fig:pentac_bg}) plus a simulated Gaussian signal 
of 435 events based on the observed production at H1.  The top plot shows the signal with a width due 
to experimental resolution $(\sigma = 4.15\:\textrm{MeV}\!\!\:/c^2)$ only 
while the bottom plot shows the signal with a width due to the 
experimental resolution plus the maximum natural width (at 95\% CL) 
based on the H1 data $(\sigma = 4.15 \oplus 16.6 = 17.1\:\textrm{MeV}\!\!\:/c^2)$.}
\label{fig:h1signal}
\end{figure}

FOCUS has published the most precise measurements of the mass and width of
four $D^{**}$ states~\cite{ddoublestar} with decay modes similar to that of the charmed
pentaquark.  However, from a FOCUS data set of 104,000 $D^{*-}$ and 137,000 $D^-$ decays,
we find no evidence for a charm pentaquark decaying strongly to $D^{(*)-}p$ over the mass 
range of 30~MeV/$c^2$ above threshold to 3.9~GeV/$c^2$.  
Results for yields and cross sections versus mass are found in 
Figs.~\ref{fig:pentac_yld} and \ref{fig:pentac_xsec}.  A summary of these limits over the
entire mass range is given in Table~\ref{tab:summary}.  The sample of $D^{*-}$ events is more than
30 times larger as well as cleaner than the sample used by H1 in the paper which found evidence
for the charm pentaquark~\cite{h1}.  While H1 finds $\sim\!1\%$ 
of the $D^{*-}$ from $\Theta_c^0$, FOCUS sets a limit of $<\!0.075\%$ at 95\% CL.  
In addition, the production is similar between the two experiments; virtual (real)
photons on protons (nucleons) for H1 (FOCUS).  Thus, the H1 result is either a statistical
fluctuation or the result of an unusual production mechanism which increases the 
charm pentaquark to charm cross section by a factor of at least 10 in H1 relative
to FOCUS\@.  A simple study to determine the extent of the difference was performed.  
\textsc{Pythia} was used
to generate $D^{*+}$ and $\Theta_c^0(3100)$ events using production appropriate for FOCUS and H1\@.  
The $\Theta_c^0(3100)$ was modeled as the average of $\Xi_c^0$ and $\Sigma_c^+$, both at 3.1~GeV/$c^2$.  
Only events in the acceptance of the experiment were accepted by requiring the pentaquark momentum be 
greater than $50\:\textrm{GeV}/c$ for FOCUS 
and requiring H1 events to have $1\!<\!Q^2\!<\!100\;\textrm{GeV}^2/c^2$ and $0.05\!<\!y\!<\!0.70$.
From these samples, the production rate of $\Theta_c^0(3100)$
to $D^{*+}$ was found to be 1.58\% (1.19\%) 
for H1 (FOCUS).  Scaling by the observed production rate of 1\% 
at H1, the production rate of $\Theta_c^0(3100)$ relative to $D^{*+}$ should be $(1.19/1.58) \cdot 0.01 = 0.75\%$ 
for FOCUS\@.  For this FOCUS analysis, the $\Theta_c^0(3100)$ efficiency relative to $D^{*+}$ is 54\%.  
Therefore, from the FOCUS sample of 107,525 $D^{*+}$ events, one would
expect $107525 \cdot 0.0075 \cdot 0.54 = 435.5$ $\Theta_c^0(3100)$ events.  
This signal has been superimposed on the FOCUS
data in Fig~\ref{fig:h1signal}.  This signal would be easily observable.
The non observation by FOCUS is also consistent with published results from 
ALEPH~\cite{nopentac_aleph} and ZEUS~\cite{nopentac_zeus}.

\begin{table}
\caption{Summary of upper limits (UL) on pentaquark yields and cross sections including systematic uncertainties.  
Results represent the maximum UL over the mass range shown in Figs.~\ref{fig:pentac_yld} and \ref{fig:pentac_xsec}.}
\label{tab:summary}
\begin{tabular}{lccc}
& Natural Width & 95\% CL UL & 95\% CL UL on \\[-5pt]
\raise8pt\hbox{Decay Mode} & (MeV/$c^2$) & on Yield   & $\sigma\left(\Theta_c^0\right) \cdot 
\textrm{BR}\left(\Theta_c^0\!\rightarrow\! D\,p\right) / \sigma\left( D \right)$ \\ \hline
                                                 & 0    & 25 & $4.2\times 10^{-4}$ \\[-6pt]
\raise10pt\hbox{$\Theta_c\!\rightarrow\!D^{*-}p$} & 16.6 & 44 & $7.5\times 10^{-4}$ \\[-3pt]
                                                 & 0    & 31 & $5.0\times 10^{-4}$ \\[-6pt]
\raise10pt\hbox{$\Theta_c\!\rightarrow\!D^-p$}    & 16.6 & 41 & $7.1\times 10^{-4}$ \\

\end{tabular}
\end{table}

\section{Acknowledgments}
We wish to acknowledge the assistance of the staffs of Fermi National
Accelerator Laboratory, the INFN of Italy, and the physics departments
of the collaborating institutions. This research was supported in part
by the U.~S.  National Science Foundation, the U.~S. Department of
Energy, the Italian Istituto Nazionale di Fisica Nucleare and
Ministero dell'Istruzione dell'Universit\`a e della Ricerca, the
Brazilian Conselho Nacional de Desenvolvimento Cient\'{\i}fico e
Tecnol\'ogico, CONACyT-M\'exico, the Korean Ministry of Education, 
and the Korean Science and Engineering Foundation.

\bibliographystyle{unsrt}

\end{document}